\documentclass{aa} 

\usepackage{amsmath}
\DeclareMathOperator{\erf}{erf}
\usepackage{float}
\makeatother
\usepackage{booktabs}
\usepackage{graphicx}
\usepackage{txfonts}
\usepackage[breaklinks=true,colorlinks=true,
            linkcolor=black,citecolor=black,urlcolor=black]{hyperref}
\usepackage{cuted}    
\usepackage{afterpage}
\usepackage{caption}    
\usepackage{booktabs}
\usepackage{graphicx}   
\usepackage{placeins}
\usepackage[dvipsnames]{xcolor}
\usepackage{needspace}
\usepackage{dblfloatfix} 
\usepackage{placeins}    
\begin{document}

   \title{How internal structure shapes the metallicity of giant exoplanets}

\author{Lorenzo Peerani\inst{1}
        \and Saburo Howard\inst{1}
        \and Ravit Helled\inst{1}
}

\institute{Institut für Astrophysik, Universität Zürich, Winterthurerstr. 190, CH8057 Zurich, Switzerland,\\
           \email{lorenzo.peerani@uzh.ch}
}

   \date{Received Month date, year; accepted Month date, year}

\abstract
{The composition and internal structure of gas giant exoplanets encode key information about their formation and evolution.}
{We investigate how different assumed interior structures affect the inferred bulk metallicity and its correlation with planetary mass.}
{For a sample of 44 giant exoplanets (0.12–5.98 $M_{J}$), we computed evolutionary models with \textsc{CEPAM} and retrieved their bulk metallicities under three structural hypotheses: core+envelope (CE), dilute core (DC), and fully mixed (FM).}
{Across all structures, we recover a significant positive correlation between total heavy-element mass ($M_{Z}$) and planetary mass ($M$), and a negative correlation between bulk metallicity ($Z$) and $M$ (also for $Z/Z_\star$ vs\ $M$). Dilute core structures yield metallicities comparable to CE models, regardless of the assumed extent of the composition gradient. Increasing atmospheric metallicity augments the inferred bulk metallicity, as enhanced opacities slow planetary cooling. Non-adiabatic DC models can further increase the retrieved metallicity by up to 35\%. We find that the mass–metallicity anti-correlation is primarily driven by low-mass, metal-rich planets ($M < 0.2\,M_{J}$), and that massive planets ($\gtrsim1\,M_{J}$) can exhibit unexpectedly high metallicities ($Z \sim 0.1$–$0.3$).}
{Improved constraints on convective mixing, combined with upcoming accurate  measurements of planetary masses, radii, and atmospheric compositions from missions such as PLATO and Ariel, will provide further constraints on  interior structure and formation models of gas giant planets.}


   \maketitle

\section{Introduction}
Inferring a planet's bulk composition and internal structure provides a powerful diagnostic of its formation pathway \citep{Helled_2011,Mordasini_2016, Turrini_2021}. Gas giant planets represent an ideal case study in this context: as the largest planets, they must have formed early, when ample gas and solids were still available for accretion. Their high masses and rapid growth mean that they preserve valuable information about the composition of the protoplanetary disk \citep{Pollack_1996, Helled_2021, Mordasini_2016}. Hence, constraining their composition is key to probing the conditions under which planets form.

The observational landscape now includes over $6{,}000$ confirmed exoplanets \citep{NASA}, with a large, well‐characterised subset of giants having measured masses and radii from space and ground-based surveys (e.g. Kepler, TESS, WASP, and HAT/HATSouth). The large dataset allows inferences of heavy‐element content in individual planets and statistical analyses across populations, enabling the identification of more general trends constrained by bulk properties.

Interior models are designed to reproduce the observed bulk properties by assuming specific distributions of heavy elements. These models often explore two simple end-member configurations: a core+envelope (CE) structure, where a distinct heavy-element core is surrounded by a H/He envelope, and a fully mixed (FM) structure, where the heavy elements are homogeneously distributed throughout the planet \citep{Thorngren_2016, Howard_2025}. However, gravity-field measurements and ring seismology for Jupiter and Saturn suggest more complex interiors. Interior models tailored to match these data suggest that the interiors of the two gas giants consist of compositional gradients of heavy elements or 'dilute (fuzzy) cores' \citep{Wahl_2017, Debras_2019, Mankovich_2021, Miguel_2022, Militzer_2022, Howard_2023, Ziv_2024}. Nevertheless, efforts in applying such a fuzzy or dilute core (DC) structure to exoplanetary studies are still embryonic. Recent retrieval studies have begun testing such structural prescriptions in individual exoplanets. For instance, \citet{Bloot_2023} applied homogeneous and inhomogeneous models within a Bayesian framework to a limited sample, highlighting strong degeneracies that prevent quantitative discrimination between the two. Similarly, \citet{vanDijk_2025} extended this approach by incorporating hot Jupiters, showing that current observations allow only qualitative assessments for a few well-characterised cases. In this study, we extend interior modelling beyond the traditional end-member cases to include composition gradients and DC structures that capture gradual heavy-element distributions.

\begin{figure*}[!t]
    \centering
    \includegraphics[width=0.9\textwidth]{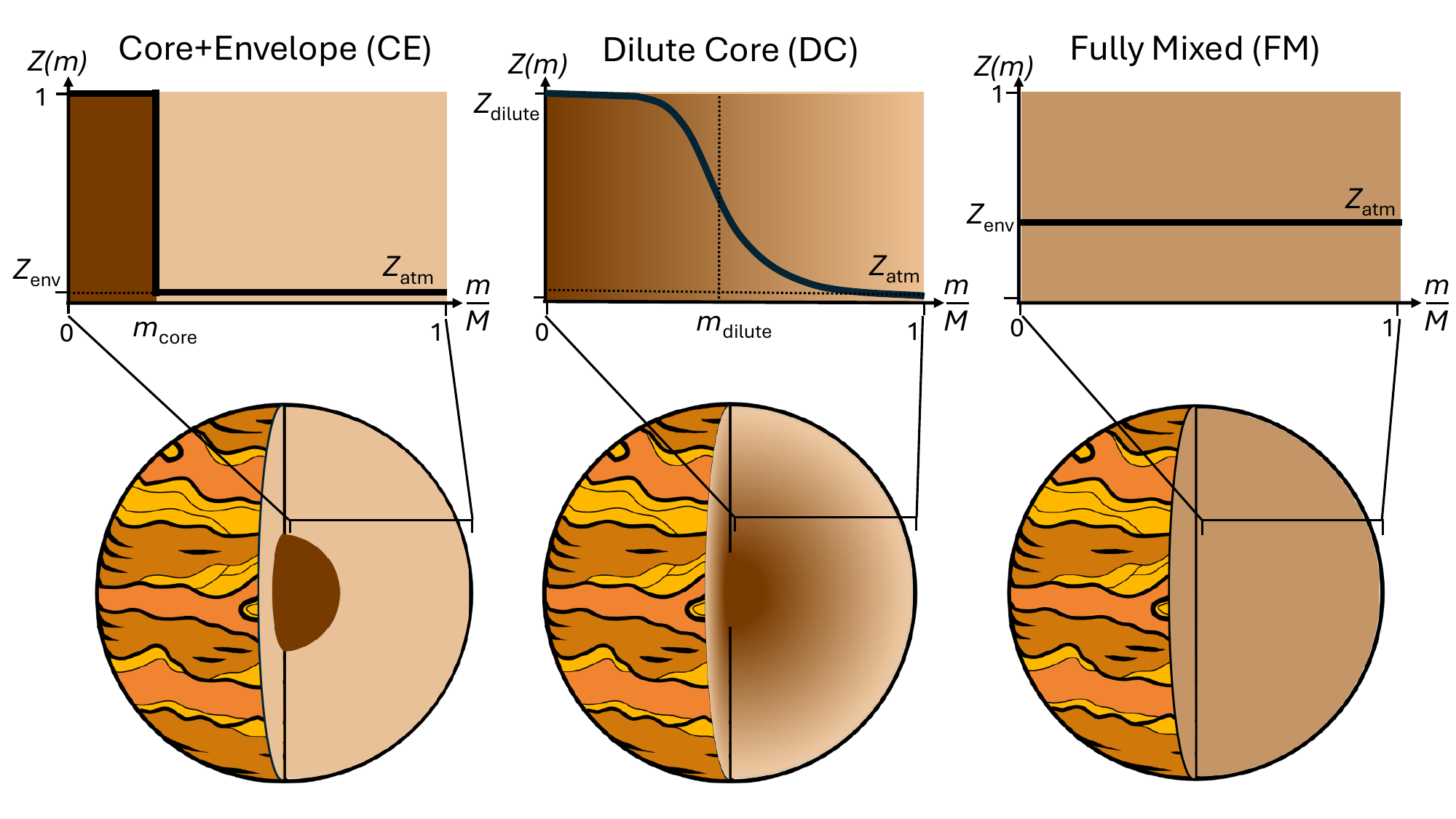} 
    \caption{Three internal structures tested in this study, with their respective heavy element structure profiles: CE, all the heavy elements situated in a well-defined core; DC, a gradient applied to the planet's core; and FM: a homogeneous envelope throughout the planet.
    }
    \label{fig.1}
\end{figure*}
In this work, we assess how the inferred bulk metallicity ($Z$) depends on the assumed interior structure for a sample of gas giant exoplanets. We further examine how different structural assumptions affect the resulting population-level trends. 
This study is important for the interpretation of JWST data  \citep{Carone_2025, Deka_2025} as well as future observations of giant planet atmospheres. Atmospheric measurements are currently being used to further constrain the planetary interiors \citep{Thorngren_2019, Wilkinson_2024,knierim2025}. Future missions such as PLATO and Ariel will provide accurate measurements of planetary radii, masses, ages, and atmospheric compositions, allowing tighter constraints on exoplanet interiors and formation histories. \citep{Rauer_2014, Tinetti_2018}.

\section{Methods}

\subsection{Model parameters}

We used the \emph{Code d’Evolution Planetaire Adaptif et Modulaire} (CEPAM) \citep{Guillot_Morel_1995, Howard_2024}  to simulate the planetary evolution. CEPAM numerically solves the 1D planetary structure and thermal-evolution equations (mass conservation, hydrostatic equilibrium, energy transport, and energy conservation) coupled to EoSs, which set density and entropy as functions of $P$, $T$, and composition. The initial catalogue of giant planets from \citet{Otegi_2020, Parc_2024} was first restricted to those with all four required observational parameters (mass, radius, age, and $T_{\rm eq}$) available. This requirement alone substantially reduced the sample size. From this subset, we applied additional filtering, requiring relative uncertainties smaller than 5\% in radius and 10\% in mass, following \citet{Howard_2025}, which yelded a final sample of 44 planets with masses between 0.12 and 5.98 $M_{\rm J}$. Exoplanets with equilibrium temperatures $T_{\rm eq}$ > 1000 K were excluded from this study, since additional energy sources can significantly inflate their radii \citep{Guillot_2006, Thorngren_2018}.

The ice-to-rock ratio was kept constant for all the tested structures and was assumed to be a 1:1 ratio. For hydrogen and helium, the ratio between the two elements was assumed to be protosolar ($Y/(X+Y)$ = 0.27) \citep{Asplund_2021}. 

 The equations of state (EoSs) used are important and are known to affect the inferred composition  \citep{Baraffe2008,Miguel_2016,Muller_2020,Howard_2025}. The EoSs used in this study are the SESAME water and dry sand for ices and rocks \citep{Lyon_1992}, and the CMS19+HG23 EoS for H-He \citep{Chabrier_2019, Howard_Guillot}. 

The atmosphere was treated as a non-grey, fully radiative atmosphere. Opacities were treated as two bands (visible and IR), and the mean opacities were tabulated accordingly \citep{Freedman_2008, Freedman_2014, Parmentier_2015}. Opacity enhancement caused by heavy elements follows the methodology described in \citet{Valencia_2013}.

For all internal structures, the $P$-$T$ profile was assumed to be adiabatic, and we did not consider compositional mixing of heavy elements. The initial entropy was not prescribed directly but instead was determined by the assumed initial planetary radius. The initial radii are typically in the range of 1–1.5 $R_{J}$, depending on the planetary mass. The choice of the initial radius is not expected to significantly influence the evolution when considering sufficiently old planets and adiabatic cooling \citep{Marley_2007, Muller_2023}. For completeness, we also considered a few cases where we assumed a non-adiabatic interior (see Section~\ref{sec:Non-adiabatic structure}).

\subsection{Tested interior structures}

We considered different interior structures for the planetary dataset. The three main tested structures include the following:
\begin{itemize}
  \item Core+envelope: This structure assumes a well-defined core consisting of heavy elements, with the envelope being composed of H-He. As the metallicity in the core $Z_{\rm core}$ is unity, we used the core mass $M_{\rm core}$ as a free parameter to infer bulk metallicities. For the basic CE scenario, there are no heavy elements in the envelope: $Z_{\rm env}$ = $Z_{\rm atm}=0$), where $Z_{\rm env}$ is the metallicity in the envelope and $Z_{\rm atm}$ in the atmosphere.  
  \item Core+envelope with stellar-abundance  atmosphere (CEA): To assess the contribution of metals mixed into the envelope, we allowed heavy elements in the upper planetary layers \(Z_{\rm env}=Z_{\rm atm}>0\). 
    \item Fully mixed: This structure assumes a homogeneous composition throughout the planetary interior. No core is present. The bulk metallicity equals to $Z_{\rm env}=Z_{\rm atm}$.

    \item Dilute core: This structure represents a gradual distribution of heavy elements. Following \citet{Miguel_2022b, Bloot_2023, Howard_2023}, we describe the structure using the following function:
    \begin{equation}
    \begin{aligned}
     Z(m)= Z_{\rm env} + \frac{Z_{\rm dilute} - Z_{\rm env}}{2}
       \left[ 1 - \erf\!\left(\frac{m - m_{\rm dilute}}{\delta m}\right) \right],
    \end{aligned}
    \label{eq.1}
    \end{equation}
    where $Z_{\rm dilute}$ is the maximum heavy–element fraction in the DC, the parameter $\delta m$ controls the steepness of the gradient, and $m_{\rm dilute}$ is the mass coordinate at which the gradient is steepest (see Fig.~\ref{fig.1}). The value $\delta m$ was fixed at 0.125 (as variations in this parameter are not believed to substantially affect planetary Z estimates \citep{Bloot_2023}). In the basic DC structure, $Z_{\rm env}=Z_{\rm atm}=0$.
    \item Dilute cores with stellar- and super-stellar abundance atmospheres (DCA and DCA3): This structure is the same as that for DC but allows for a non-zero atmospheric metallicity. The atmospheric abundances tested match the host star's ($Z_{\star}$) metal abundance ($Z_{\rm atm}=Z_{\star}$). Similar to Jupiter’s atmospheric enrichment \citep{Helled_2024b}, a more enriched case with $Z_{\rm atm}=3\, \times Z_\star$ has also been tested (DCA3). Note that in these cases, $m_{\rm dilute}$ was fixed to a value of 0.5.
\end{itemize}
For each structure that is consistent with the observational constraints for a given planet, we report $M_{Z}$ (heavy-element mass) and $Z$ (bulk metallicity), related by $Z = M_{Z}/M$, where $M$ is the planetary mass.

\subsection{Root-find $Z$ retrieval}
\label{sec:rootfind}
In this section, we describe the statistical tool designed to retrieve the possible ranges of $Z$ that reproduce each planetary radius for every structural assumption. The retrieval of planetary bulk metallicities $Z$ was designed as an inversion of the
evolution model: instead of using $M$ and $Z$ to predict $(R, \text{Age})$, we seek the metallicity that reproduces the observed $(M, R, \text{Age})$ within their uncertainties. Our approach incorporates observational uncertainties by treating the mass and radius as normally distributed around their observed values
($M \sim \mathcal{N}(M_{\text{obs}}, \sigma_{M})$ and ($R \sim \mathcal{N}(R_{\text{obs}}, \sigma_{R})$) , while the age is assumed to follow a uniform distribution   
($\text{Age} \sim \mathcal{U}(\text{Age}_{\min}, \text{Age}_{\max})$). The atmospheric metallicity was not used as an additional constraint, since this measurement is not available for the majority of the planets in our sample. The method proceeds in three steps:

\begin{itemize}
    \item Step one – Evolution runs:
    We first generated a set of cooling tracks with CEPAM across a broad grid of input parameters.  
    Metallicities were sampled in the range $0 \leq Z \leq 1$ with steps of 0.025, while masses were varied around the observed mean value, and sampled at the central value, $\pm 1\sigma$, $\pm 2\sigma$, and $\pm 3\sigma$.
    This ensemble of runs ensures that the parameter space captures both the observational uncertainties in mass and the possible degeneracies 
    between metallicity and mass in shaping planetary radii.  
    A dense coverage of $Z$ values is particularly important for detecting degeneracies in the $M$–$Z$–$R$ relation.

    \item Step two – Grid interpolation:
    The model outputs from CEPAM (describable as a radius/age relation for different structures and chemical abundances) were interpolated to build a continuous parameter grid.  
    While linear interpolation is generally sufficient, we employed cubic spline interpolation 
    (via \texttt{scipy.interpolate.interp1d} \citep{Virtanen_2020}) since it better resolves curved behaviours and possible degeneracies in the cooling tracks.  
    The interpolated grid was refined to a resolution of $\Delta Z = 0.0025$, corresponding to 400 metallicity steps\footnote{The choice of step sizes reflects a balance between 
    accuracy and computational time. Tests on two representative planets with different $M$–$Z$ step combinations showed that denser sampling in 
    metallicity is more critical than in mass. These steps are sufficiently small for our dataset. Future data with better observational constraints, or simply applying such a grid to Jupiter and Saturn, might render this sampling too coarse and unreliable and would require adjustments.}. This step provides smooth predictions of planetary radii across the 3D space $(M, Z, \text{Age})$, 
    which are later required for the inversion procedure.
    \item Step three – Root–finding and sampling:
    \begin{figure*}[!htbp]
    \centering
    \includegraphics[width=\textwidth]{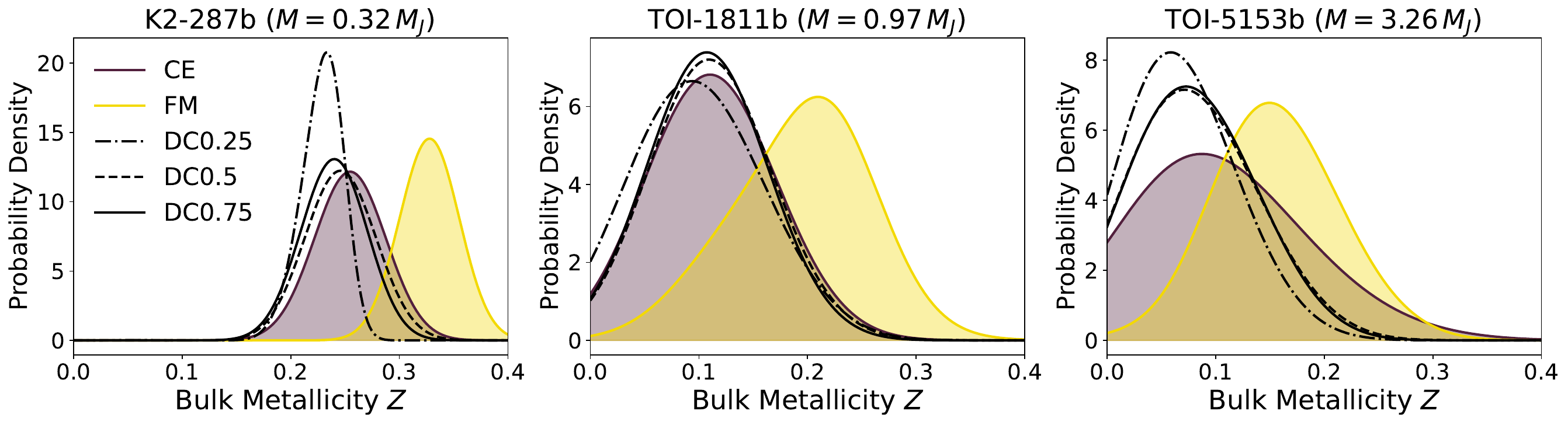} 
    \caption{Retrieved curves for CE, FM, and DC0.25-0.5-0.75 structures (varying $m_{\rm dilute}$ values) across three different planetary cases for different masses. The CE curves are shown in purple, FM in yellow, and DC in black, and can be distinguished through the different line style.
    }
    \label{fig.2}
\end{figure*} 
    To invert the grid and estimate $Z$, we repeatedly sampled observational values of mass, radius, and age.  
    For each planet, we drew $n=5000$ tuples $(M,R, \text{Age})$.  
    For each tuple, Brent’s method was applied to solve:

\begin{equation}
\begin{aligned}    
    f(Z) = R_{\text{pred}}(Z,M,\text{Age}) - R_{\text{obs}} = 0 ,
    \end{aligned}
    \label{eq.2}
    \end{equation}
    
    where $R_{\text{pred}}$ is the interpolated model prediction and $R_{\text{obs}}$ is the sampled observational value.  
    This procedure yields the metallicity $Z$ that reproduces the observed radius.  
    Repeating this process $n$ times provides a posterior distribution for a prior of $Z \sim \mathcal{U}(0, 1)$, capturing both measurement uncertainties and intrinsic model degeneracies. A similar approach was implemented in \citet{Muller_2023}.  
\end{itemize}
Root-finding retrieval provides the posterior distributions of metallicity $Z$ for each assumed interior structure and for every exoplanet analysed. These distributions were smoothed using a kernel density estimate (KDE), and the corresponding mean and standard deviation of $Z$ were extracted and used to investigate planetary trends.

 \begin{figure*}[!b]
    \centering
    \includegraphics[width=\textwidth]{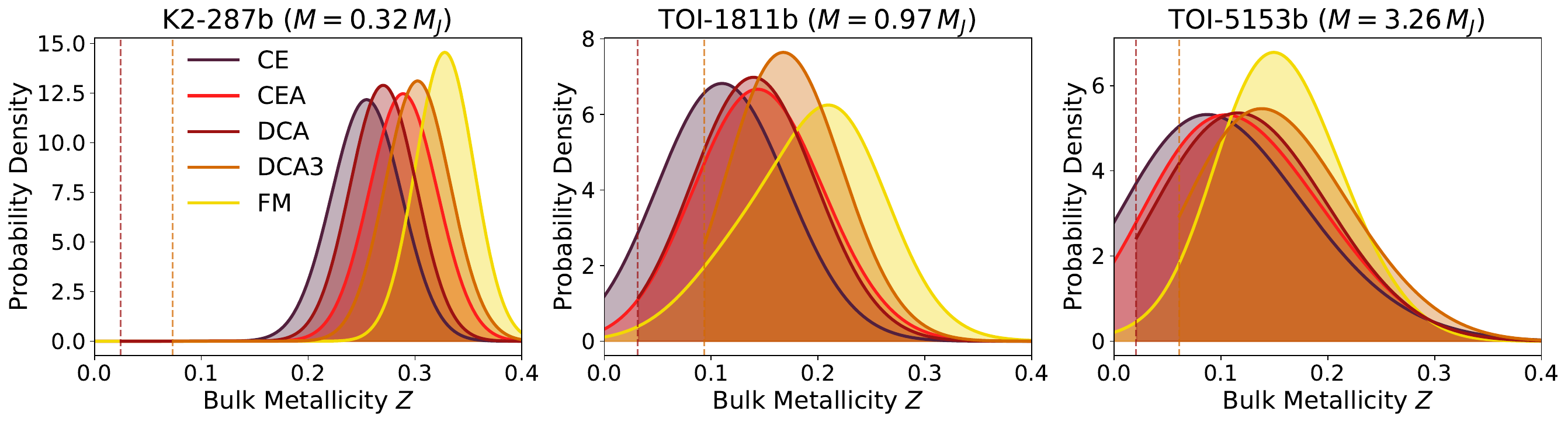} 
    \caption{Retrieved curves for CE, FM, CEA, DCA, and DCA3. The dashed lines (in burgundy and orange) show the atmospheric abundances assumed for DCA and DCA3 models, respectively. Such lines act as a lower threshold for these models as the assumed $Z_{\rm atm}$ was kept fixed for a given planet.
    }
    \label{fig.3}
\end{figure*}

\section{Results}
\subsection{Structural properties}
Figure~\ref{fig.2} shows the retrieved curves for CE, FM, and three DC structures that assume a different extent of the DC ($m_{\rm dilute}=$ 0.25, 0.5 or 0.75). Since $m_{\rm dilute}$ remained fixed in the three different cases, {$Z_{\rm dilute}$} was used as the adjusting parameter. This setup allowed us to explore the impact of changing the extent of the dilution curve. Although the analysis was performed for the full sample, we first focus on three planets with representative masses: K2-287b (0.315~$M_{J}$), TOI-1811b (0.972~$M_{J}$), and TOI-5153b (3.26~$M_{J}$). In all cases, CE yields a lower bulk metallicity than FM. We verified this offset by reproducing the approach of \citet{Howard_2025}; the results are consistent with our pipeline (see Appendix~\ref{app:howard25}). The physical driver is that, in FM, heavy elements mixed into the envelope and atmosphere increase the mean molecular weight and, critically, the Rosseland mean opacity in the radiative zone. The higher opacity slows cooling and keeps radii larger at fixed age; matching the observed radius then requires higher heavy-element mass than in the CE case. \citet{Howard_2025} also found that FM evolution tracks are systematically warmer, especially at younger ages. The opacity effect is the main contributor in FM structures for higher expected bulk metallicities.

When examining the DC curves for different $m_{\rm dilute}$ values (Fig.~\ref{fig.2}), we find that DC models generally have lower values of $Z$ compared to CE and FM models, and their solutions more closely resemble CE than FM. No systematic distinction or trend is evident across the three different $m_{\rm dilute}$ values tested, as the model-to-model differences (and their intrinsic uncertainties) remain smaller than the observational errors. Since these DC structures exclude atmospheric metallicity ($Z_{\rm atm}=0$), the observed effects are driven solely by the internal distribution of heavy elements. Introducing a gradient, rather than a sharp core boundary, redistributes heavy elements into outer layers, increasing the local density and pressure. This results in a lower radius over time, because a smaller heavy-element mass is required to reproduce the observed radius and age. Nonetheless, such an effect appears to be of lower magnitude compared to the difference in FM-CE estimates.

After testing different $m_{\rm dilute}$ structures, we next consider DC models with atmospheric metallicities set to one and three times the stellar value (DCA and DCA3, respectively). The results for the three representative planets are shown in Fig.~\ref{fig.3}. The complete dataset, including mean values and standard deviations for all structures, is presented in Appendix~\ref{app:planetary_data}. Since $Z_{\rm atm}$ (and thus, the opacity effect) is the varying parameter, DCA and DCA3 directly probe its effect on the $Z$ estimates. As shown in Fig.~\ref{fig.3}, increasing the atmospheric metallicity systematically increases the inferred $Z$.

In the three cases shown, both DCA and DCA3 yield higher $Z$ than CE. For some planets (such as HAT-P-12b, TOI-2010b, and TOI-2180b), however, DCA results converge with DC models, suggesting that the opacity effects are minor. Overall, Fig.~\ref{fig.3} confirms that the atmospheric opacity significantly affects the retrieved heavy-element abundance, whereas density redistribution alone produces comparatively minor shifts. The CEA (bright-red) curves closely track the DCA estimates (especially for TOI-1811b and TOI-5153b), indicating that both models have similar $Z$. Because they use identical opacities and differ only in interior structure (distinct core in CEA vs.~composition gradient in DCA), this agreement reinforces that opacity is the dominant lever over the density distribution.

However, while we find that FM models generally lead to higher inferred bulk metallicities, we note that DCA3 models can sometimes lead to even higher estimates, especially for low-$Z$ planets (for example in the cases of TOI-4127b and TOI-2589b). Furthermore, the results are affected by the choice of $Z_{\rm atm}$ since our method does not allow for solutions where the inferred bulk metallicity is lower than the assumed atmospheric metallicity. This limitation particularly affects low-$Z$ planets, as illustrated by the cutoff in the DCA and DCA3 posterior distribution of TOI-5153b shown in Fig.~\ref{fig.3}.
Knowledge of the planetary atmospheric composition is therefore crucial for narrowing the parameter space of possible solutions.

\subsection{Mass-metallicity Relation}

After analysing the overall behaviour of the metallicity estimates across different structural models, 
we next examined the full dataset of 44 planets for possible population-level trends in the inferred 
composition. In particular, we investigated the heavy-element mass ($M_{Z}$), and the bulk metallicity ($Z$), and their dependency on the planetary mass ($M$). 
\begin{figure}[!hbp]
    \centering
    \includegraphics[width=0.5\textwidth,trim={2mm 2mm 2mm 2mm},clip]{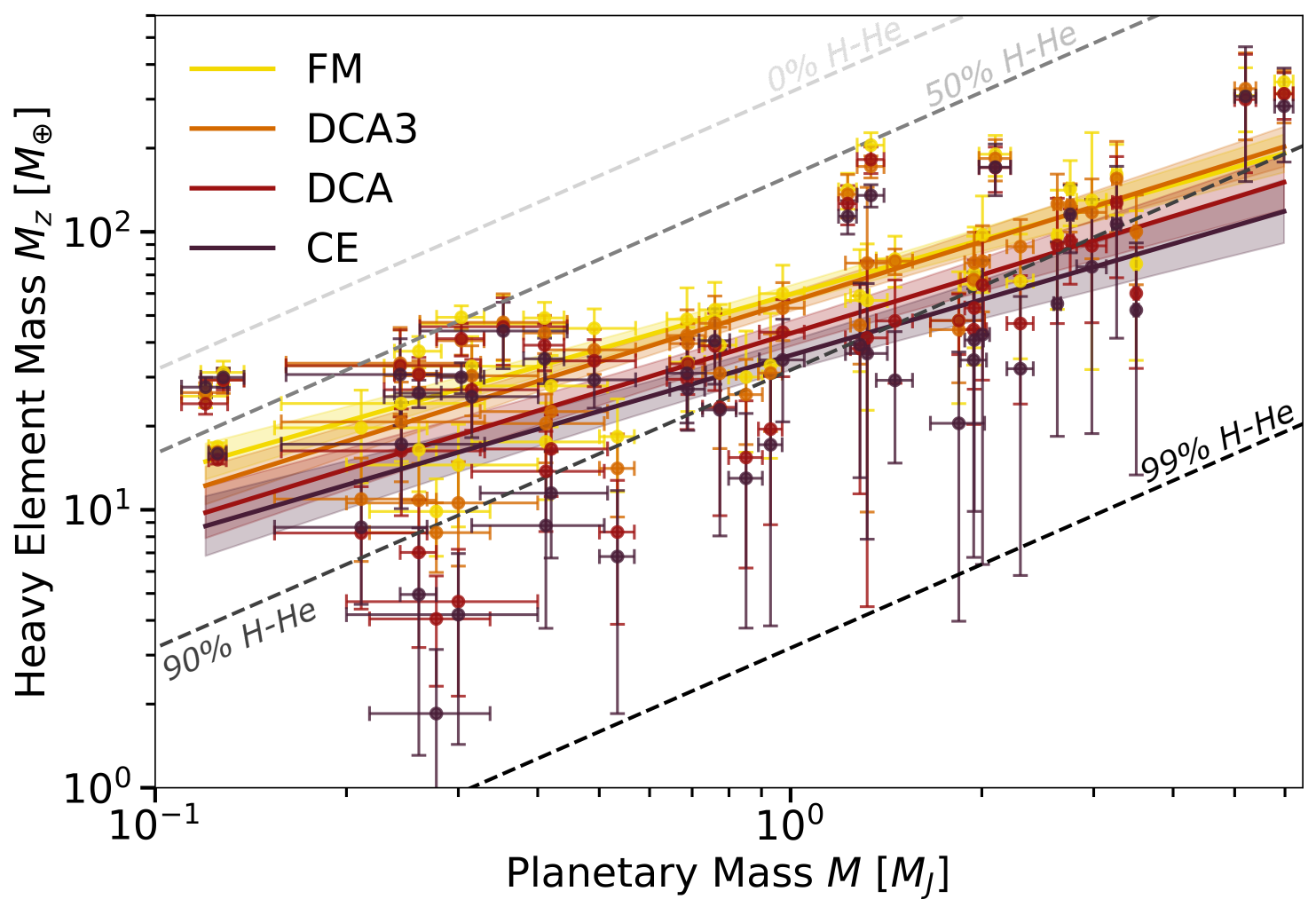}
    \caption{Relation between the retrieved heavy-element mass $M_{Z}$ and planetary mass $M$, shown for the four main structures tested (CE, FM, DCA, and DCA3). Shown as dashed lines in a grey gradient are curves indicating the trends for fixed planetary compositions (0\%, 50\%, 90\%, and 99\% gas composition, respectively). Bayesian fits are shown using a power-law function $M_{Z} = \beta \, M^{\alpha}$. A summary of the fitted parameters and correlation coefficients is presented in Table~\ref{tab:alphabeta} and Table~\ref{tab:tau}.}
    \label{fig.4}
\end{figure}
To assess these relations, we performed a Bayesian regression. The parameters of the power law $M_{Z}\,[M_{\oplus}] = \beta \, M\,[M_{J}]^{\alpha}$ were estimated using a Markov chain Monte Carlo (MCMC, Python package \texttt{emcee} \citep{Foreman_2013}, following \citet{Muller_2023} and \citet{Howard_2025}). Posterior distributions of the slope ($\alpha$) and intercept ($\beta$) 
were sampled from 5,000 chains, taking into account $1\sigma$ uncertainties in both axes.The outputs for each tested structure is reported in Table~\ref{tab:alphabeta}.

All models reveal a positive slope between $M_{Z}$ and $M$, with similar slopes but different intercepts depending on the assumed structure (see Fig.~\ref{fig.4}). The FM model systematically predicts the highest $M_{Z}$, 
while CE predicts the lowest (with $\beta \approx 36$). This scaling is broadly consistent with the 
relation reported by \citet{Thorngren_2016}, \citet{Howard_2025}, and \citet{Chachan_2025}. Our results suggest a steeper dependence, with DCA3 yielding $\alpha > 0.7 \pm 0.07$, only marginally inconsistent with Thorngren's relation of $M_{Z} \propto \sqrt{M}$ when uncertainties are considered.

For this reason, we inspected the reliability of such curves by testing the correlation 
between $\alpha$ and $\beta$. The covariance analysis shows a weak but positive coupling 
($\rho \sim 0.2$ for all the models), indicating that the slope and intercept are not fully independent, and that small 
shifts in one parameter affect the other. We then also evaluated the fits with a weighted 
least-squares (WLS) $\chi^2$ test. In all cases, $\chi^2 \gg 1$, suggesting that the power-law 
relations do not formally explain the data within the quoted errors; therefore, the reliance on such curves should be treated with caution. Not meeting such reliability limits with the $\chi^2$ statistic could be caused by (i) noise arising from the scatter in the data, (ii) a large diversity in planetary compositions, or both. In Section~\ref{app:wsl}, we further discuss the WSL goodness-of-fit approach.

\begin{figure}[!t]
    \centering
    \includegraphics[width=0.5\textwidth]{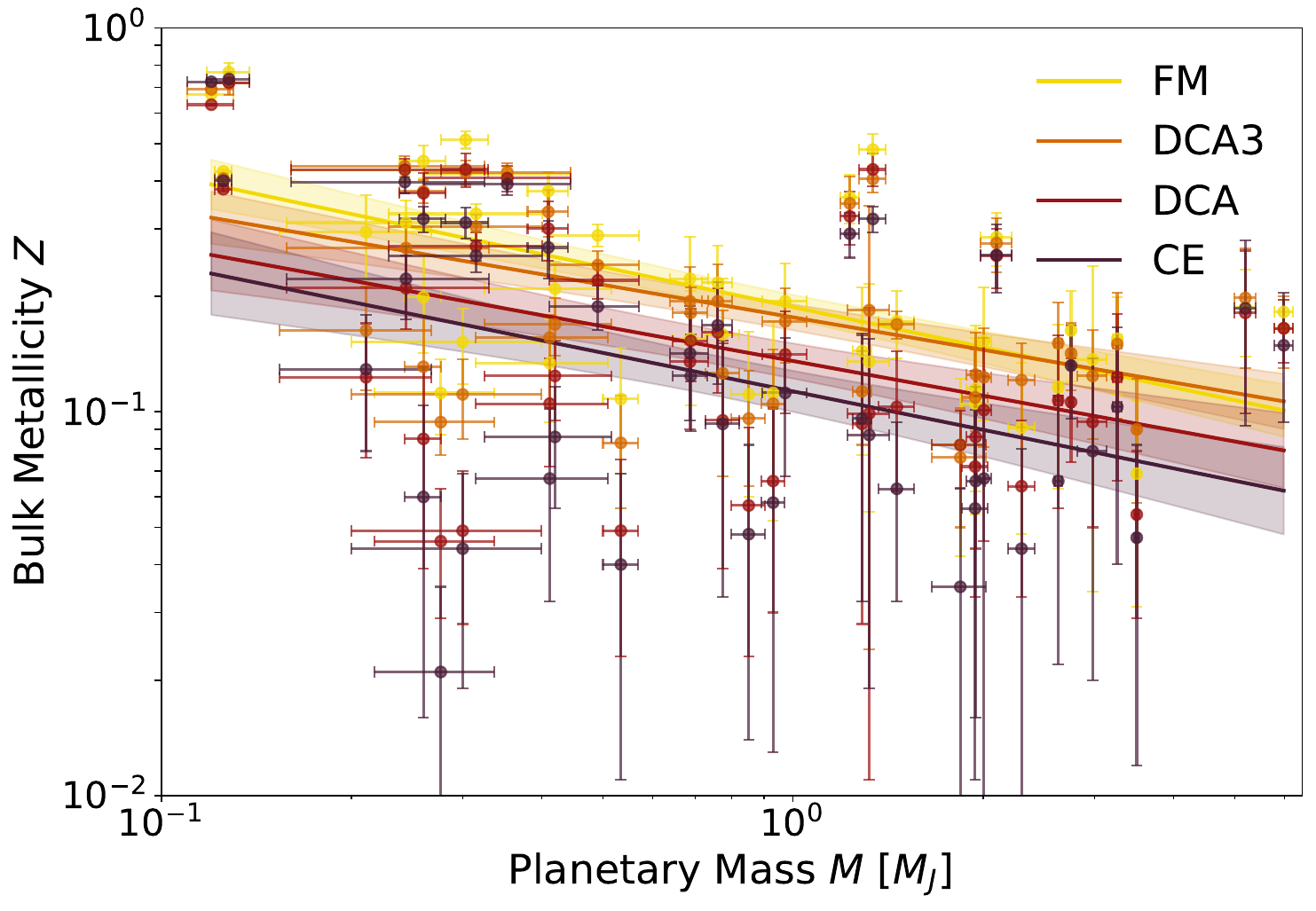} 
    \caption{Relation between retrieved metallicity $Z$ and planetary mass $M$, shown for the four main structures tested (CE, FM, DCA, and DCA3). Bayesian fits are shown using a power-law function, $Z = \beta \, M^{\alpha}$. For additional quantitative results, we refer to Tables~\ref{tab:alphabeta} and~\ref{tab:tau}.}
    \label{fig.5}
\end{figure}

Since the WSL test revealed the limited reliability of the retrieved power-law fits, it is essential to verify whether the apparent slopes of these relations genuinely reflect an underlying correlation within the data.
We further tested the correlations of the dataset for each structural assumption using the non-parametric Kendall’s $\tau$ rank test. 
To account for observational uncertainties, instead of computing the correlation on the mean values of the data, a bootstrap resampling approach was employed. Here, 5,000 synthetic datasets were generated by drawing from the reported error distributions, and $\tau$ was calculated for each realisation. The median $\tau$ values and associated 
$p$-values are listed in Table~\ref{tab:tau}. For the $M_{Z}$--$M$ relation, nearly all runs 
returned $p$-values well below the adopted significance threshold of $p=0.05$, confirming the 
robustness of the positive correlation.

\begin{table}[ht]
  \caption[Power-law fit parameters for $M_Z$–$M$ and $Z$–$M$]{%
    Power-law fit parameters ($y = \beta\,M^{\alpha}$) derived from Bayesian regressions
    for the four interior-structure models.}
  \label{tab:alphabeta}
  \centering
  \small
  \setlength{\tabcolsep}{4pt}
  \begin{tabular}{l|cc|cc}
    \toprule
    \textbf{Model} &
    \multicolumn{2}{c|}{\textbf{(a) $M_Z$–$M$}} &
    \multicolumn{2}{c}{\textbf{(b) $Z$–$M$}} \\
    & $\boldsymbol{\alpha}$ &
      $\boldsymbol{\beta}$ &
      $\boldsymbol{\alpha}$ &
      $\boldsymbol{\beta}$ \\
    \midrule
    CE   & 0.67 $\pm$ 0.11 & 35.9 $\pm$ 4.5 & $-0.33$ $\pm$ 0.12 & 0.113 $\pm$ 0.014 \\
    DCA  & 0.70 $\pm$ 0.10 & 43.1 $\pm$ 4.4 & $-0.30$ $\pm$ 0.10 & 0.136 $\pm$ 0.015 \\
    DCA3 & 0.72 $\pm$ 0.07 & 56.0 $\pm$ 4.2 & $-0.28$ $\pm$ 0.07 & 0.176 $\pm$ 0.014 \\
    FM   & 0.65 $\pm$ 0.07 & 59.6 $\pm$ 4.4 & $-0.35$ $\pm$ 0.07 & 0.187 $\pm$ 0.014 \\
    \bottomrule
  \end{tabular}
  \tablefoot{%
  Parameters $\alpha$ and $\beta$ correspond to the best-fit power-law relations
  $M_Z=\beta\,M^{\alpha}$ and $Z=\beta\,M^{\alpha}$. 
  Quoted values are posterior medians from Bayesian fits; typical $1\sigma$ uncertainties are
  $\sigma_{\alpha}\!\simeq\!0.07$ and $\sigma_{\beta}\!\simeq\!4.5$ for $M_Z$–$M$, and
  $\sigma_{\alpha}\!\simeq\!0.07$, $\sigma_{\beta}\!\simeq\!0.014$ for $Z$–$M$.}
\end{table}

\begin{table}[ht]
\caption[Kendall’s $\tau$ and p-values across models]{%
  Kendall’s $\tau$ coefficients and associated $p$-values for four structural models
  (CE, DCA, DCA3, and FM) across the two tested relations, (a) $M_Z$–$M$ and (b) $Z$–$M$.}
  \label{tab:tau}
  \centering
  \small
  \setlength{\tabcolsep}{10pt}
  \begin{tabular}{l|cc|cc}
    \toprule
    \textbf{Model} &
    \multicolumn{2}{c|}{\textbf{(a) $M_Z$–$M$}} &
    \multicolumn{2}{c}{\textbf{(b) $Z$–$M$}} \\
    & $\boldsymbol{\tau}$ & \textbf{$p$-value} &
      $\boldsymbol{\tau}$ & \textbf{$p$-value} \\
    \midrule
    CE   & 0.514 & $1.2\times10^{-6}$  & $-0.273$ & $1.0\times10^{-2}$ \\
    DCA  & 0.612 & $7.6\times10^{-9}$  & $-0.245$ & $2.1\times10^{-2}$ \\
    DCA3 & 0.660 & $4.4\times10^{-10}$ & $-0.314$ & $3.1\times10^{-3}$ \\
    FM   & 0.660 & $4.4\times10^{-10}$ & $-0.397$ & $1.8\times10^{-4}$ \\
    \bottomrule
  \end{tabular}
\end{table}

Next, we examined the relation between $Z$ and $M$ (Fig.~\ref{fig.5}). 
In contrast to the $M_{Z}$--$M$ trend, the $Z$--$M$ relation shows a clear negative correlation. 
The ranking of models remains similar: FM predicts the highest average $Z$, followed by DCA3, DCA, 
and CE. Compared to the $M_{Z}$--$M$ relation, the data are more scattered, particularly 
at lower planetary masses. Kendall’s $\tau$ again indicates statistically significant negative correlations, although the correlation strength is weaker. 
A portion of the reduced significance may reflect mathematical coupling from dividing by the $x$–axis variable ($Z=M_{Z}/M$), which weakens the apparent correlation and inflates $p$-values.
Median $p$-values remain generally 
below 0.05, but with a modest fraction of realisations exceeding the threshold 
($\sim 2\%$ for FM and DCA3, and $\sim 10$--$11\%$ for DCA and CE). 
Nevertheless, the correlations remain statistically compelling across all models.
\begin{figure}[!b]
  \centering
  \includegraphics[width=0.5\textwidth]{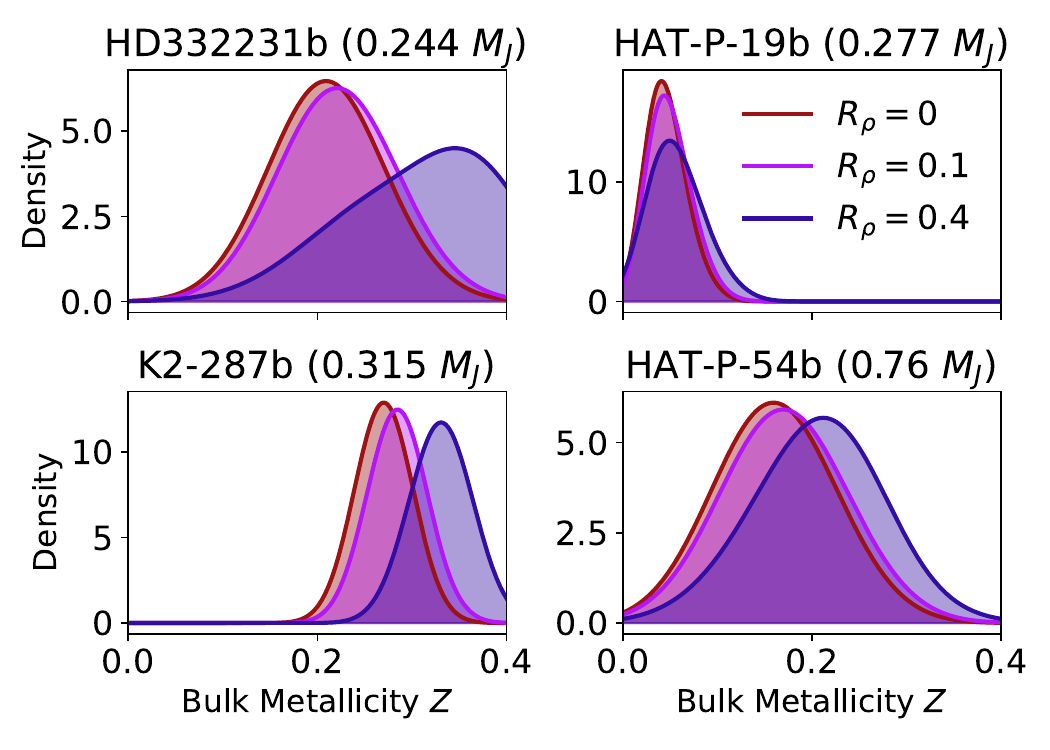}
  \vspace{4mm} 
  \includegraphics[width=0.5\textwidth]{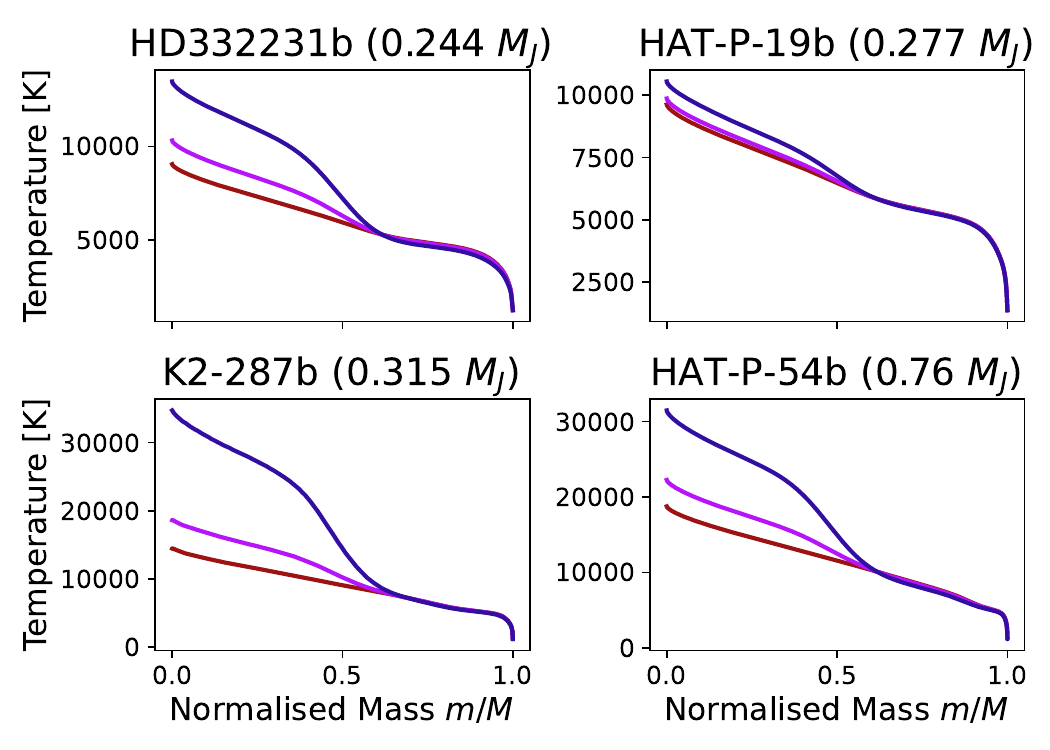}
  \caption{Top: Metallicity estimates for adiabatic ($R_\rho$ = 0) and super-adiabatic ($R_\rho$ = 0.1, 0.4) models. This is shown for four different planetary cases. Bottom: Corresponding temperature profiles over the planet's normalised mass, using the structures from the median metallicity of the DCA model at a fixed age of 4~Gyr.
  }
  \label{fig.6}
\end{figure}

A similar relationship was tested using the normalisation to the host star ($Z$/$Z_{\star}$ vs $M$). The results show similar trends (both in best-fit and correlation tests) to those obtained for $Z$/$M$. Such an investigation would have been essential if we had artificially imposed an anti-correlation driven by the stellar metallicity (especially since one of the model assumptions is that the stellar composition reflects that the planet's atmospheric structure). Although recent studies \citep{Buchhave_2014, Muller_2024a} have found trends between planetary heavy-element mass and host star chemical composition, the dataset used in this study does not show any correlation between $Z_{\star}$ and $M$.

\section{Discussion}
\begin{figure*}[!b]
    \centering
    \includegraphics[width=0.9\textwidth]{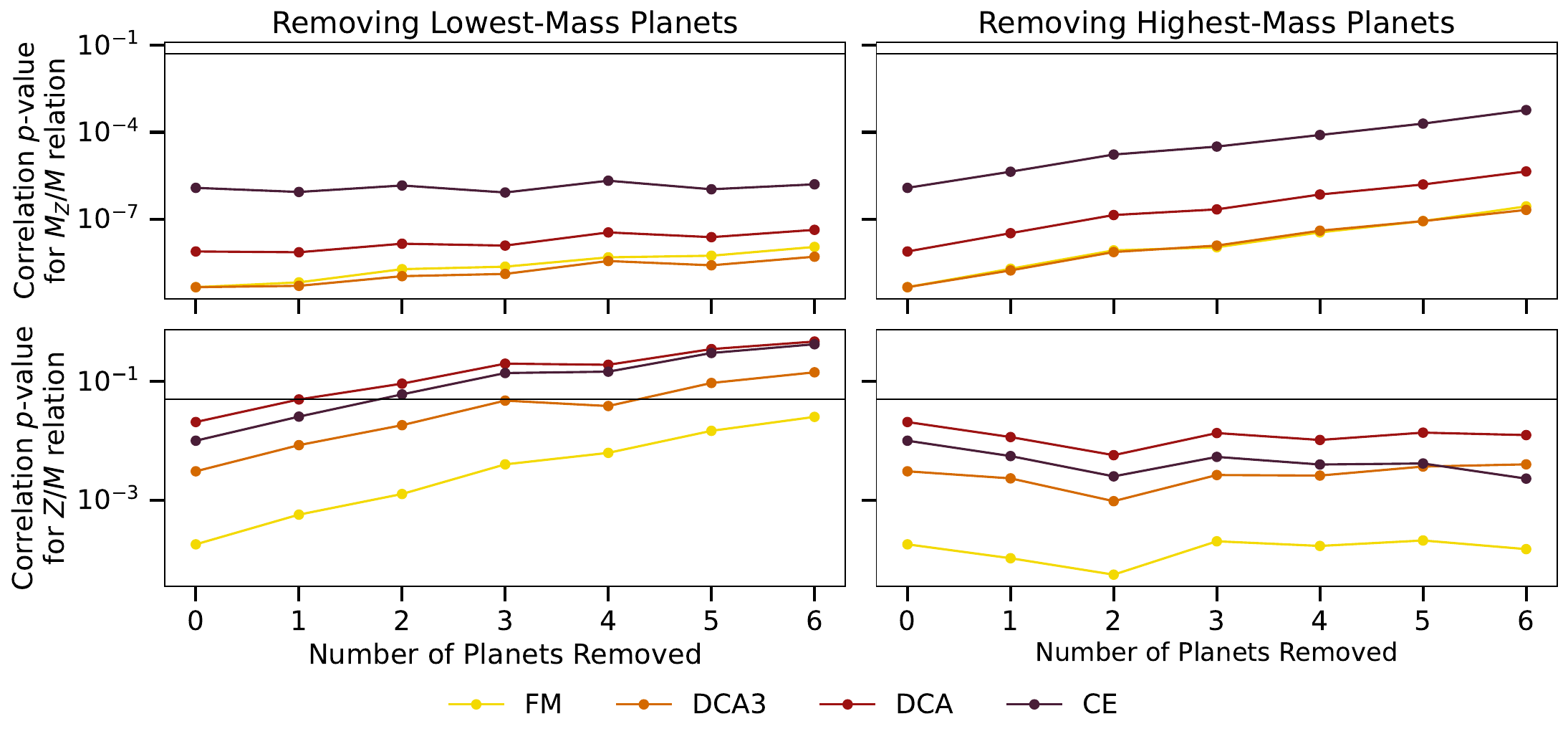}
    \caption{Kendall's $\tau$ correlation p-value for $M_Z/M$ and $Z/M$ with $n$-planets removed from the initial dataset. Planets are removed from the lowest-mass (left) and highest-mass (right) planets. The number of planets being iteratively removed can be seen on the x-axis. The four different models are respectively colour-coded. The horizontal line shows the $p=0.05$ acceptance threshold.}
    \label{fig.7}
\end{figure*}

\subsection{Non-adiabatic structure}
\label{sec:Non-adiabatic structure}
In our models with a DC (DC, DCA, and DCA3), we assumed an adiabatic temperature gradient. However, the presence of a composition gradient implies that the temperature gradient could differ from the adiabatic one. To assess the effect of a non-adiabatic temperature gradient on the inferred bulk metallicity, we parameterised the deviation from adiabaticity (likely caused by double-diffusive convection \citep{Debras_2021, Leconte_2012}) as in \citet{Howard2025_demixing}, using
\begin{equation}
    \nabla_{T}=\nabla_{\rm ad}+R_{\rho}B,
\label{eq.3}
\end{equation}
where $B = \frac{\chi_{\rho}}{\chi_{\rm T}}\left(\frac{d\,\textrm{ln}\,\rho}{d\, \textrm{ln}\,Z} \right)_{P,T} \nabla_{Z}$, $\chi_{\rho} = \left(\frac{d\,\textrm{ln}\,P}{d\, \textrm{ln}\,\rho} \right)_{T,Z}$, $\chi_{\rm T} = \left(\frac{d\,\textrm{ln}\,P}{d\, \textrm{ln}\,T} \right)_{\rho,Z}$ and $\nabla_{Z}=\frac{d\,\textrm{ln}\,Z}{d\,\textrm{ln}\,P}$.

For a smaller selection of planets, we simulated the planetary evolution using $R_{\rho}=0.1$ and 0.4 and compared it to the adiabatic case ($R_{\rho}=0$). 
The inferred bulk metallicities $Z$ are shown in Fig.~\ref{fig.6}. We find that $R_{\rho}=0.1$ and $R_{\rho}=0.4$ lead to increased values of $Z$ by up to 7~\% and 35~\%, respectively. The super-adiabatic temperature gradients obtained with $R_{\rho}=0.1$ and $R_{\rho}=0.4$ yield hotter interiors compared to the adiabatic case (see bottom panel of Fig.~\ref{fig.6}) and naturally lead to an increased amount of heavy elements required to match the planets' radii.
We note that efforts to constrain $R_{\rho}$ for the Solar System’s gas giants have been recently presented \citep{Sur_2025}, although the strong degeneracy in this parameter resulted in a broad range of admissible values. Previous studies estimated   $R_{\rho}$ for Jupiter and Saturn, suggesting $R_{\rho}$<0.1 at around 0.05 \citep{Mankovich_2016, Mankovich_2020}. This led us to test values of 0.1 and an extreme case of 0.4. The actual value is model-dependent, and efforts to constrain it in giant planets still remains ongoing.

A clear limitation of our models is that convective mixing is not included. This process could affect the thermal evolution and the distribution of heavy elements. For example, \citet{knierim2025} showed that for a Jupiter-mass planet neglecting mixing could overestimate the radius by 0.08~$R_{J}$. The values of $R_{\rho}$ we chose were arbitrary, and it is necessary to assess whether DCs with such a temperature structure can remain stable against convection. Future studies should further assess the effect of non-adiabaticity on the mass--metallicity relation and the thermal evolution of gas giants.

\subsection{Connection with formation}
\paragraph{Extent of the DC.}
For some high-metallicity cases (\(Z\gtrsim 0.5\); e.g., TOI-4010d and WASP-196b), no solutions were found for compact gradients with $m_{\rm dilute}=0.25$ or 0.5: the interior could not accommodate the required heavy-element fraction unless the gradient was shifted outwards to higher $m_{\rm dilute}$. Although this behaviour might suggest a more extended dilute region in these planets, this interpretation is not uniquely supported by the data. The inferred extent of dilution is strongly model-dependent and can arise simply from our assumption of a fixed $Z_{\rm atm}$, which permits additional heavy elements to be stored in the envelope without altering the internal composition gradient.
In more massive planets, a wide non-adiabatic DC could substantially increase the heat retained in the deep interior, altering the radius evolution. Constraining $m_{\rm dilute}$ is therefore crucial for massive planets. A well-confined DC may have little impact on inflation (the thermal profile remains close to adiabatic), whereas high $m_{\rm dilute}$ values can stabilise significantly more heat. In non-adiabatic cases, the steepness of the gradient may also become important. 

\paragraph{Mass-metallicity relation.}
Our results, obtained across different structural configurations, are consistent with previous studies \citep{Thorngren_2016, Teske_2019, Howard_2025, Chachan_2025}. We recover a clear positive correlation between the total heavy-element mass ($M_{Z}$) and planetary mass ($M$), and a weaker, yet significant, negative correlation between metallicity ($Z$) and $M$, also when normalised by $Z_\star$.

The overall $Z$–$M$ trend indicates a gradual decrease in planetary metallicity with increasing mass. There is a sharp decrease in $Z$ for low planetary masses (0.2-0.3 $M_{J}$), followed by a flatter behaviour for higher masses. Other studies on mass-metallicity relations have found similar results \citep{Chachan_2025}.
At the high-mass end, we find hints of renewed enrichment.  
This enrichment, together with the significant dispersion across the mass–metallicity relation, likely reflects the diversity of formation pathways and interior structures of giant planets. The presence of high-$Z$, low-$M$ objects (potentially with compositions more akin to Neptune-like planets) is consistent with planet formation theory \citep{Helled2023}. Interestingly, we find that more massive planets (M > 4 $M_{J}$) are also metal-rich. Even if discerning any conclusion from two planets is premature, this may suggest a secondary accretion of heavy elements through late planetesimal capture or giant impacts after disk dispersal, leading to enhanced metal enrichment in their deep envelopes \citep{Mordasini_2016, Helled_2017, Venturini_2020,ShibataHelled}.

To assess the robustness of the inferred anti-correlation, we examined how the correlation significance changes when progressively removing planets from the mass extremes. We quantified this sensitivity by recalculating the Kendall $\tau$ correlation and the corresponding $p$-values after each removal for both the $M_{Z}/M$ and $Z/M$ relations (Fig.~\ref{fig.7}). 

We find that removing low-mass planets has a small effect on the inferred $M_{Z}/M$ relation (top row in Fig.~\ref{fig.7}): the $p$-values remain nearly unchanged for all the models. In contrast, removing high-mass objects systematically increases the $p$-values. These constitute the 'anchor points' that most strongly support the monotonic trend found with the correlation test. Even after the removal of six planets from either end, the correlation remains statistically significant, confirming that the observed trend is robust.

For the $Z/M$ relation, the behaviour is different. Because $Z$ already scales inversely with mass ($Z = M_{Z}/M$), plotting  $Z$ against $M$ effectively introduces an additional mass dependence in the denominator, which leads to a higher inferred $p$-value \citep{Kronmal_1993, Brett_2004}. The removal of two low-mass planets is sufficient for the CE and DCA models to lose significance at the $p = 0.05$ level. After the removal of six planets, only the FM model retains a marginally significant $p$-value. In contrast, removing the most massive planets has a rather small impact. We find that removing high-mass planets from the sample increases the $p$-values for $M_{Z}$ versus $M$, while removing low-mass planets increases the $p$-values for $Z$ versus $M$. This is because $Z = M_{Z}/M$.

Our findings raise a classification question. Planets with $Z \gtrsim 0.5$ and low masses likely reside near the transition between Neptunian and Jovian regimes, where electron degeneracy pressure no longer dominates the $M$--$R$ relation \citep{Muller_2024b}. 
The contrast in how the $p$-values increase when removing either low-$M$ or high-$M$ planets could hold a physical meaning. This may indicate the diversity of the sample and the potential existence of sub-populations that do not follow a monotonic trend. 
In this context, the assumption of a universal, continuous mass--metallicity trend is clearly oversimplified and cannot reflect the large diversity in planetary composition.

\section{Conclusions}
We quantified bulk metallicities for transiting gas giants under multiple interior structures using \textsc{CEPAM} with a root–finding retrieval. Our key conclusions can be summarised as follows:

\begin{itemize} 
\item We confirm that the inferred bulk metallicity for FM structures is higher than for CE structures, due to the opacity effect that prevails over the density redistribution effect. This, in turn, leads to DC structures that yield similar bulk metallicities to CE, independently of the extent of the composition gradient. 
\item When increasing the atmospheric metallicity in the DC structure to one (DCA) or three times stellar (DCA3), the inferred bulk metallicity becomes closer to the one yielded by FM. This demonstrates the importance of calibrating it with future observational data to better constrain internal metallicity estimates.
\item  When considering non-adiabatic temperature gradients in DC structures, the inferred bulk metallicity can increase by up to 35\% for a used value of $R_{\rho} = 0.4$. However, future studies should explore a larger parameter space and assess the long-term stability of composition gradients.
\item  Using a sample of 44 giant exoplanets, we confirm that heavy-element mass increases with planetary mass. The slope of the $M_Z - M$ relationship does not change with the assumed structure, but only the intercept.
\item  The planetary bulk metallicity $Z$ generally decreases with planetary mass. Planets with $M$ < 0.2 M$_J$ have metallicities between 0.4 and 0.8. These low M -- high Z planets are required to hold the negative correlation between Z and M. Planets with 0.2 < Mp < 4~$M_{J}$ have a strong scatter in metallicities, which can range from less than 0.01 to 0.5, showing the diverse composition of giant exoplanets. For the two high planetary masses ($M$ > 4$M_{J}$) in our sample, the inferred metallicities remain relatively high (between 0.1 and 0.3). This suggests that giant planets continue to accrete heavy elements after their formation \citep{ShibataHelled}. 
\end{itemize}

Future studies that account for additional physical processes such as convective mixing as well as the observed atmospheric compositions can improve giant planet characterisation. In addition, further investigations of breakpoints via multi-linear or orthogonal-distance regressions \citep{Muller_2024b} in the inferred relations are desirable.
Accurate measurements of the atmospheric chemistry of giant planets by JWST can also further constrain the planetary bulk composition and internal structure. In addition, forthcoming measurements from PLATO and Ariel (expected to be launched in 2026 and 2029, respectively) will sharpen bulk metallicity estimates by providing tighter constraints on mass, radius, age, and atmospheric composition for many planets. This will allow a better understanding of the interiors of giant planets and of their formation and evolution histories.

\begin{acknowledgements}
We thank Stefano Wirth, Simon Müller, Henrik Knierim, and Shay Zucker for insightful discussions.
\end{acknowledgements}

\bibliographystyle{aa}
\bibliography{biblio} 

\clearpage
\begin{appendix}

\section{Planetary data}
\label{app:planetary_data}

\begin{strip}
\captionof{table}{Exoplanet data used in this study.}
\label{tab:sample_fullwidth_1}

\vspace{0.5em}

\begin{center}
\begin{tabular*}{\textwidth}{@{\extracolsep{\fill}}lcccccccc}
\toprule
Planet & M ($M_J$) & R ($R_J$) & Age (Myr) & $Z_{st}$ & T (K) & $Z_{\rm CE}$ & $Z_{\rm FM}$ \\
\midrule
TOI-4010d & 0.12 $\pm$ 0.01 & 0.551 $\pm$ 0.013 & 3000–9200 & 0.036 $\pm$ 0.07 & 650 & 0.724 $\pm$ 0.011 & 0.672 $\pm$ 0.026 \\
HATS-72b & 0.125 $\pm$ 0.004 & 0.722 $\pm$ 0.003 & 11710–12410 & 0.019 $\pm$ 0.014 & 739 & 0.4 $\pm$ 0.008 & 0.423 $\pm$ 0.004 \\
WASP-156b & 0.128 $\pm$ 0.01 & 0.51 $\pm$ 0.02 & 2400–10400 & 0.027 $\pm$ 0.12 & 970 & 0.736 $\pm$ 0.01 & 0.767 $\pm$ 0.045 \\
HAT-P-12b* & 0.211 $\pm$ 0.057 & 0.959 $\pm$ 0.025 & 500–4500 & 0.008 $\pm$ 0.05 & 963 & 0.129 $\pm$ 0.05 & 0.294 $\pm$ 0.073 \\
TOI-3629b* & 0.243 $\pm$ 0.082 & 0.74 $\pm$ 0.014 & 5000–13800 & 0.055 $\pm$ 0.093 & 711 & 0.397 $\pm$ 0.026 & 0.427 $\pm$ 0.025 \\
HD332231b* & 0.244 $\pm$ 0.086 & 0.867 $\pm$ 0.027 & 2400–6800 & 0.017 $\pm$ 0.059 & 876 & 0.222 $\pm$ 0.048 & 0.311 $\pm$ 0.045 \\
K2-329b & 0.26 $\pm$ 0.022 & 0.774 $\pm$ 0.026 & 500–4000 & 0.019 $\pm$ 0.068 & 650 & 0.318 $\pm$ 0.024 & 0.45 $\pm$ 0.044 \\
WASP-69b & 0.26 $\pm$ 0.017 & 1.057 $\pm$ 0.047 & 500–3000 & 0.021 $\pm$ 0.077 & 963 & 0.06 $\pm$ 0.044 & 0.199 $\pm$ 0.057 \\
HAT-P-19b* & 0.277 $\pm$ 0.06 & 1.008 $\pm$ 0.014 & 3200–11200 & 0.023 $\pm$ 0.07 & 981 & 0.021 $\pm$ 0.014 & 0.112 $\pm$ 0.025 \\
TOI-4406b* & 0.3 $\pm$ 0.1 & 1.0 $\pm$ 0.02 & 2200–3600 & 0.019 $\pm$ 0.05 & 904 & 0.044 $\pm$ 0.025 & 0.152 $\pm$ 0.034 \\
TOI-1268b & 0.303 $\pm$ 0.026 & 0.812 $\pm$ 0.054 & 110–380 & 0.035 $\pm$ 0.06 & 919 & 0.311 $\pm$ 0.029 & 0.511 $\pm$ 0.026 \\
K2-287b* & 0.315 $\pm$ 0.086 & 0.847 $\pm$ 0.013 & 3700–5700 & 0.024 $\pm$ 0.04 & 804 & 0.255 $\pm$ 0.024 & 0.328 $\pm$ 0.02 \\
HATS-49b* & 0.353 $\pm$ 0.092 & 0.765 $\pm$ 0.013 & 8500–11900 & 0.025 $\pm$ 0.053 & 835 & 0.392 $\pm$ 0.025 & 0.416 $\pm$ 0.018 \\
WASP-132b & 0.41 $\pm$ 0.03 & 0.87 $\pm$ 0.03 & 500–2500 & 0.026 $\pm$ 0.13 & 763 & 0.268 $\pm$ 0.046 & 0.376 $\pm$ 0.044 \\
TOI-5344b* & 0.412 $\pm$ 0.097 & 0.946 $\pm$ 0.021 & 5100–13800 & 0.041 $\pm$ 0.088 & 689 & 0.067 $\pm$ 0.035 & 0.134 $\pm$ 0.04 \\
TOI-201b* & 0.42 $\pm$ 0.095 & 1.008 $\pm$ 0.014 & 380–1330 & 0.027 $\pm$ 0.036 & 759 & 0.086 $\pm$ 0.03 & 0.209 $\pm$ 0.037 \\
HATS-75b* & 0.491 $\pm$ 0.079 & 0.884 $\pm$ 0.013 & 10600–13800 & 0.051 $\pm$ 0.4 & 772 & 0.188 $\pm$ 0.025 & 0.288 $\pm$ 0.019 \\
HAT-P-17b* & 0.534 $\pm$ 0.034 & 1.01 $\pm$ 0.029 & 4500–11100 & 0.015 $\pm$ 0.08 & 792 & 0.04 $\pm$ 0.029 & 0.108 $\pm$ 0.039 \\
WASP-84b* & 0.687 $\pm$ 0.048 & 0.976 $\pm$ 0.025 & 500–3700 & 0.019 $\pm$ 0.12 & 833 & 0.142 $\pm$ 0.047 & 0.222 $\pm$ 0.064 \\
TOI-3714b* & 0.689 $\pm$ 0.044 & 0.944 $\pm$ 0.02 & 6000–13800 & 0.038 $\pm$ 0.086 & 764 & 0.124 $\pm$ 0.035 & 0.155 $\pm$ 0.051 \\
HAT-P-54b* & 0.76 $\pm$ 0.042 & 0.944 $\pm$ 0.028 & 1800–8200 & 0.011 $\pm$ 0.08 & 818 & 0.168 $\pm$ 0.05 & 0.217 $\pm$ 0.054 \\
K2-290c & 0.774 $\pm$ 0.047 & 1.006 $\pm$ 0.05 & 3200–5600 & 0.013 $\pm$ 0.1 & 676 & 0.093 $\pm$ 0.06 & 0.159 $\pm$ 0.063 \\
TOI-1478b & 0.851 $\pm$ 0.052 & 1.06 $\pm$ 0.04 & 5200–12200 & 0.018 $\pm$ 0.069 & 918 & 0.048 $\pm$ 0.034 & 0.111 $\pm$ 0.051 \\
K2-140b & 0.93 $\pm$ 0.04 & 1.21 $\pm$ 0.09 & 5300–13200 & 0.019 $\pm$ 0.1 & 962 & 0.058 $\pm$ 0.045 & 0.112 $\pm$ 0.06 \\
TOI-1811b* & 0.972 $\pm$ 0.079 & 0.994 $\pm$ 0.025 & 1900–10800 & 0.031 $\pm$ 77.0 & 962 & 0.112 $\pm$ 0.044 & 0.194 $\pm$ 0.049 \\
WASP-130b & 1.23 $\pm$ 0.04 & 0.89 $\pm$ 0.03 & 400–7900 & 0.028 $\pm$ 0.1 & 833 & 0.291 $\pm$ 0.039 & 0.362 $\pm$ 0.052 \\
TOI-2010b* & 1.286 $\pm$ 0.044 & 1.054 $\pm$ 0.027 & 600–4200 & 0.023 $\pm$ 0.055 & 400 & 0.096 $\pm$ 0.064 & 0.144 $\pm$ 0.067 \\
TOI-5542b & 1.32 $\pm$ 0.1 & 1.009 $\pm$ 0.036 & 7200–12900 & 0.009 $\pm$ 0.08 & 441 & 0.087 $\pm$ 0.068 & 0.135 $\pm$ 0.08 \\
HATS-17b & 1.338 $\pm$ 0.065 & 0.777 $\pm$ 0.056 & 800–3400 & 0.031 $\pm$ 0.03 & 814 & 0.318 $\pm$ 0.025 & 0.482 $\pm$ 0.047 \\
HATS-74Ab* & 1.46 $\pm$ 0.096 & 1.032 $\pm$ 0.021 & 5900–13800 & 0.05 $\pm$ 0.027 & 895 & 0.063 $\pm$ 0.031 & 0.172 $\pm$ 0.034 \\
Kepler-117c & 1.84 $\pm$ 0.18 & 1.101 $\pm$ 0.035 & 3900–6700 & 0.014 $\pm$ 0.1 & 704 & 0.035 $\pm$ 0.028 & 0.082 $\pm$ 0.04 \\
TIC237913194b & 1.942 $\pm$ 0.092 & 1.117 $\pm$ 0.051 & 4000–7400 & 0.021 $\pm$ 0.05 & 974 & 0.056 $\pm$ 0.045 & 0.105 $\pm$ 0.051 \\
HAT-P-15b & 1.946 $\pm$ 0.066 & 1.072 $\pm$ 0.043 & 5200–9300 & 0.026 $\pm$ 0.08 & 904 & 0.066 $\pm$ 0.05 & 0.115 $\pm$ 0.053 \\
TOI-4515b & 2.005 $\pm$ 0.052 & 1.086 $\pm$ 0.039 & 1000–1400 & 0.017 $\pm$ 0.03 & 705 & 0.067 $\pm$ 0.057 & 0.153 $\pm$ 0.058 \\
K2-114b & 2.1 $\pm$ 0.12 & 0.932 $\pm$ 0.031 & 2700–11700 & 0.04 $\pm$ 0.037 & 701 & 0.256 $\pm$ 0.052 & 0.285 $\pm$ 0.045 \\
TOI-4127b & 2.3 $\pm$ 0.11 & 1.096 $\pm$ 0.036 & 2700–6900 & 0.021 $\pm$ 0.12 & 605 & 0.044 $\pm$ 0.036 & 0.091 $\pm$ 0.043 \\
HATS-76b* & 2.629 $\pm$ 0.034 & 1.079 $\pm$ 0.031 & 600–13300 & 0.032 $\pm$ 0.057 & 940 & 0.066 $\pm$ 0.044 & 0.116 $\pm$ 0.053 \\
TOI-2180b* & 2.755 $\pm$ 0.03 & 1.01 $\pm$ 0.021 & 6800–9600 & 0.028 $\pm$ 0.057 & 348 & 0.132 $\pm$ 0.036 & 0.163 $\pm$ 0.043 \\
NGTS-20b & 2.98 $\pm$ 0.16 & 1.07 $\pm$ 0.04 & 1400–6800 & 0.022 $\pm$ 0.08 & 688 & 0.079 $\pm$ 0.059 & 0.137 $\pm$ 0.103 \\
TOI-5153b* & 3.26 $\pm$ 0.05 & 1.06 $\pm$ 0.04 & 4400–6400 & 0.02 $\pm$ 0.08 & 906 & 0.103 $\pm$ 0.063 & 0.155 $\pm$ 0.044 \\
TOI-2589b* & 3.5 $\pm$ 0.029 & 1.08 $\pm$ 0.03 & 9000–13000 & 0.02 $\pm$ 0.04 & 582 & 0.047 $\pm$ 0.035 & 0.069 $\pm$ 0.038 \\
WASP-162b & 5.2 $\pm$ 0.2 & 1.0 $\pm$ 0.05 & 10620–13800 & 0.029 $\pm$ 0.13 & 910 & 0.186 $\pm$ 0.094 & 0.187 $\pm$ 0.048 \\
TOI-2338b & 5.98 $\pm$ 0.2 & 1.0 $\pm$ 0.02 & 5000–9000 & 0.026 $\pm$ 0.04 & 799 & 0.149 $\pm$ 0.055 & 0.182 $\pm$ 0.014 \\
\bottomrule
\end{tabular*}
\end{center}

\vspace{-0.5em}
\caption*{\footnotesize
\textbf{Notes.} Planetary parameters are taken from \citet{Howard_2025} and are available on \texttt{dace.unige.ch}. In addition to the planetary parameters, the retrieved bulk metallicity means and standard deviation for the CE and FM are reported on the rightmost columns. Planet names with the asterisk (*) are making up the initial 21 planets that have been tested with tighter constraints on the radius uncertainties (smaller than 3\% [$\sigma_R<3\%$] instead of 5\% used for the full dataset.
}

\clearpage
\captionof{table}{List of bulk metallicity estimates found by the study.}
\label{tab:sample_fullwidth_2}

\vspace{0.5em}

\begin{center}
\begin{tabular*}{\textwidth}{@{\extracolsep{\fill}}lcccccccc}
\toprule
Planet Name & Z$_{\mathrm{DC0.25}}$ & Z$_{\mathrm{DC0.5}}$ & Z$_{\mathrm{DC0.75}}$ & Z$_{\mathrm{DCA}}$ & Z$_{\mathrm{DCA3}}$ & Z$_{\mathrm{CEA}}$ \\
\midrule
TOI-4010d &  &  &  & 0.631 $\pm$ 0.002 & 0.692 $\pm$ 0.012 &  \\
HATS-72b &  &  &  & 0.380 $\pm$ 0.006 & 0.405 $\pm$ 0.005 &  \\
WASP-156b &  &  &  & 0.720 $\pm$ 0.004 & 0.720 $\pm$ 0.050 &  \\
HAT-P-12b* & 0.119 $\pm$ 0.048 & 0.124 $\pm$ 0.047 & 0.124 $\pm$ 0.047 & 0.123 $\pm$ 0.047 & 0.163 $\pm$ 0.049 & 0.129 $\pm$ 0.050 \\
TOI-3629b* &  & 0.389 $\pm$ 0.028 & 0.372 $\pm$ 0.027 & 0.427 $\pm$ 0.029 & 0.436 $\pm$ 0.027 & 0.453 $\pm$ 0.026 \\
HD332231b* & 0.195 $\pm$ 0.035 & 0.214 $\pm$ 0.045 & 0.209 $\pm$ 0.043 & 0.210 $\pm$ 0.046 & 0.267 $\pm$ 0.045 & 0.222 $\pm$ 0.049 \\
K2-329b &  &  &  & 0.371 $\pm$ 0.047 & 0.375 $\pm$ 0.026 &  \\
WASP-69b &  &  &  & 0.085 $\pm$ 0.046 & 0.131 $\pm$ 0.046 &  \\
HAT-P-19b* & 0.019 $\pm$ 0.014 & 0.021 $\pm$ 0.014 & 0.021 $\pm$ 0.015 & 0.046 $\pm$ 0.017 & 0.094 $\pm$ 0.017 & 0.048 $\pm$ 0.017 \\
TOI-4406b* & 0.038 $\pm$ 0.025 & 0.043 $\pm$ 0.025 & 0.043 $\pm$ 0.024 & 0.049 $\pm$ 0.021 & 0.111 $\pm$ 0.026 & 0.051 $\pm$ 0.021 \\
TOI-1268b &  &  &  & 0.428 $\pm$ 0.044 & 0.421 $\pm$ 0.031 &  \\
K2-287b* & 0.229 $\pm$ 0.015 & 0.246 $\pm$ 0.024 & 0.240 $\pm$ 0.023 & 0.270 $\pm$ 0.023 & 0.303 $\pm$ 0.023 & 0.289 $\pm$ 0.024 \\
HATS-49b* &  & 0.385 $\pm$ 0.027 & 0.367 $\pm$ 0.024 & 0.406 $\pm$ 0.032 & 0.421 $\pm$ 0.024 & 0.419 $\pm$ 0.025 \\
WASP-132b &  &  &  & 0.300 $\pm$ 0.053 & 0.332 $\pm$ 0.046 &  \\
TOI-5344b* & 0.064 $\pm$ 0.034 & 0.065 $\pm$ 0.033 & 0.064 $\pm$ 0.034 & 0.105 $\pm$ 0.033 & 0.156 $\pm$ 0.023 & 0.108 $\pm$ 0.035 \\
TOI-201b* & 0.053 $\pm$ 0.031 & 0.084 $\pm$ 0.029 & 0.082 $\pm$ 0.029 & 0.124 $\pm$ 0.029 & 0.169 $\pm$ 0.029 & 0.130 $\pm$ 0.031 \\
HATS-75b* & 0.184 $\pm$ 0.026 & 0.182 $\pm$ 0.025 & 0.179 $\pm$ 0.024 & 0.220 $\pm$ 0.023 & 0.241 $\pm$ 0.021 & 0.233 $\pm$ 0.026 \\
HAT-P-17b* & 0.041 $\pm$ 0.028 & 0.038 $\pm$ 0.028 & 0.040 $\pm$ 0.028 & 0.049 $\pm$ 0.026 & 0.083 $\pm$ 0.027 & 0.050 $\pm$ 0.026 \\
WASP-84b* & 0.124 $\pm$ 0.051 & 0.137 $\pm$ 0.046 & 0.135 $\pm$ 0.044 & 0.135 $\pm$ 0.045 & 0.194 $\pm$ 0.044 & 0.140 $\pm$ 0.048 \\
TOI-3714b* & 0.121 $\pm$ 0.038 & 0.121 $\pm$ 0.035 & 0.120 $\pm$ 0.035 & 0.153 $\pm$ 0.033 & 0.181 $\pm$ 0.030 & 0.161 $\pm$ 0.036 \\
HAT-P-54b* & 0.154 $\pm$ 0.048 & 0.162 $\pm$ 0.049 & 0.158 $\pm$ 0.048 & 0.161 $\pm$ 0.049 & 0.194 $\pm$ 0.048 & 0.166 $\pm$ 0.051 \\
K2-290c &  &  &  & 0.095 $\pm$ 0.056 & 0.126 $\pm$ 0.058 &  \\
TOI-1478b &  &  &  & 0.057 $\pm$ 0.034 & 0.096 $\pm$ 0.032 &  \\
K2-140b &  &  &  & 0.066 $\pm$ 0.036 & 0.105 $\pm$ 0.040 &  \\
TOI-1811b* & 0.099 $\pm$ 0.046 & 0.110 $\pm$ 0.042 & 0.108 $\pm$ 0.041 & 0.141 $\pm$ 0.042 & 0.172 $\pm$ 0.038 & 0.145 $\pm$ 0.044 \\
WASP-130b &  &  &  & 0.323 $\pm$ 0.051 & 0.349 $\pm$ 0.061 &  \\
TOI-2010b* & 0.037 $\pm$ 0.028 & 0.040 $\pm$ 0.031 & 0.062 $\pm$ 0.038 & 0.093 $\pm$ 0.065 & 0.113 $\pm$ 0.031 & 0.074 $\pm$ 0.031 \\
TOI-5542b &  &  &  & 0.099 $\pm$ 0.088 & 0.184 $\pm$ 0.160 &  \\
HATS-17b &  &  &  & 0.428 $\pm$ 0.042 & 0.404 $\pm$ 0.031 &  \\
HATS-74Ab* & 0.056 $\pm$ 0.030 & 0.064 $\pm$ 0.031 & 0.063 $\pm$ 0.030 & 0.103 $\pm$ 0.041 & 0.169 $\pm$ 0.014 & 0.106 $\pm$ 0.031 \\
Kepler-117c &  &  &  & 0.082 $\pm$ 0.018 & 0.076 $\pm$ 0.026 &  \\
TIC237913194b &  &  &  & 0.072 $\pm$ 0.039 & 0.109 $\pm$ 0.036 &  \\
HAT-P-15b &  &  &  & 0.086 $\pm$ 0.042 & 0.125 $\pm$ 0.036 &  \\
TOI-4515b &  &  &  & 0.101 $\pm$ 0.055 & 0.123 $\pm$ 0.042 &  \\
K2-114b &  &  &  & 0.255 $\pm$ 0.045 & 0.275 $\pm$ 0.044 &  \\
TOI-4127b &  &  &  & 0.064 $\pm$ 0.031 & 0.121 $\pm$ 0.030 &  \\
HATS-76b* & 0.046 $\pm$ 0.030 & 0.028 $\pm$ 0.014 & 0.060 $\pm$ 0.038 & 0.107 $\pm$ 0.051 & 0.151 $\pm$ 0.042 & 0.097 $\pm$ 0.046 \\
TOI-2180b* & 0.096 $\pm$ 0.029 & 0.105 $\pm$ 0.030 & 0.098 $\pm$ 0.030 & 0.106 $\pm$ 0.032 & 0.142 $\pm$ 0.028 & 0.142 $\pm$ 0.062 \\
NGTS-20b &  &  &  & 0.094 $\pm$ 0.044 & 0.124 $\pm$ 0.039 &  \\
TOI-5153b* & 0.067 $\pm$ 0.040 & 0.080 $\pm$ 0.045 & 0.079 $\pm$ 0.044 & 0.123 $\pm$ 0.057 & 0.150 $\pm$ 0.054 & 0.118 $\pm$ 0.058 \\
TOI-2589b* & 0.034 $\pm$ 0.023 & 0.038 $\pm$ 0.026 & 0.038 $\pm$ 0.026 & 0.054 $\pm$ 0.025 & 0.090 $\pm$ 0.032 & 0.065 $\pm$ 0.033 \\
WASP-162b &  &  &  & 0.181 $\pm$ 0.082 & 0.198 $\pm$ 0.068 &  \\
TOI-2338b &  &  &  & 0.165 $\pm$ 0.031 & 0.165 $\pm$ 0.035 &  \\
\bottomrule
\end{tabular*}
\end{center}

\vspace{-0.5em} 
\caption*{\footnotesize
\textbf{Notes.} Retrieved bulk metallicities for the following tested models: DC0.25, DC0.5, DC0.75, DCA, DCA3, and CEA. Planet names without the asterisk (*) are missing data from DC and CEA models; hence, the statistical analysis
was carried out only for the 21 planets with radius uncertainties smaller than 3\% ($\sigma_R<3\%$).
}
\end{strip}

\thispagestyle{empty}
\null         
\clearpage 

\begin{table}[t]  
  \centering
  \setlength{\tabcolsep}{3pt}
  \caption[List of planetary parameters]{%
    List of retrieved adiabatic (DCA) and super-adiabatic runs for D structures (tested \texorpdfstring{$R_{\rho}$} values of 0.1 and 0.4) for 11 planets.}
  \label{tab:superadiabatic_samples}

  \resizebox{\columnwidth}{!}{%
    \begin{tabular}{lccc}
      \toprule
      Planet & $R_{\rho}=0$ (DCA) & $R_{\rho}=0.1$ & $R_{\rho}=0.4$ \\
      \midrule
      HAT-P-12b   & 0.123 $\pm$ 0.047 & 0.1232 $\pm$ 0.043 & 0.189 $\pm$ 0.047 \\
      TOI-3629b   & 0.427 $\pm$ 0.029 & 0.4309 $\pm$ 0.0284 & 0.4752 $\pm$ 0.0249 \\
      HD332231b   & 0.210 $\pm$ 0.046 & 0.2222 $\pm$ 0.047 & 0.3127 $\pm$ 0.0651 \\
      HAT-P-19b   & 0.046 $\pm$ 0.017 & 0.0492 $\pm$ 0.022 & 0.055 $\pm$ 0.022 \\
      K2-287b     & 0.270 $\pm$ 0.023 & 0.2845 $\pm$ 0.0237 & 0.3306 $\pm$ 0.0252 \\
      HATS-49b    & 0.406 $\pm$ 0.032 & 0.4266 $\pm$ 0.0444 & 0.4342 $\pm$ 0.0194 \\
      HATS-75b    & 0.220 $\pm$ 0.023 & 0.2201 $\pm$ 0.0253 & 0.2735 $\pm$ 0.0287 \\
      HAT-P-17b   & 0.049 $\pm$ 0.026 & 0.050 $\pm$ 0.025 & 0.063 $\pm$ 0.036 \\
      WASP-84b    & 0.135 $\pm$ 0.045 & 0.1448 $\pm$ 0.0464 & 0.155 $\pm$ 0.045 \\
      HAT-P-54b   & 0.161 $\pm$ 0.049 & 0.1708 $\pm$ 0.0502 & 0.2054 $\pm$ 0.0529 \\
      TOI-1811b   & 0.141 $\pm$ 0.042 & 0.142 $\pm$ 0.042 & 0.177 $\pm$ 0.044 \\
      \bottomrule
    \end{tabular}
  }
\end{table}

\par\bigskip
\Needspace{6\baselineskip}   
\section{Result comparison with Howard et al. (2025)}
\label{app:howard25}

\begin{figure}[!b]
    \centering
    \includegraphics[width=1\columnwidth]{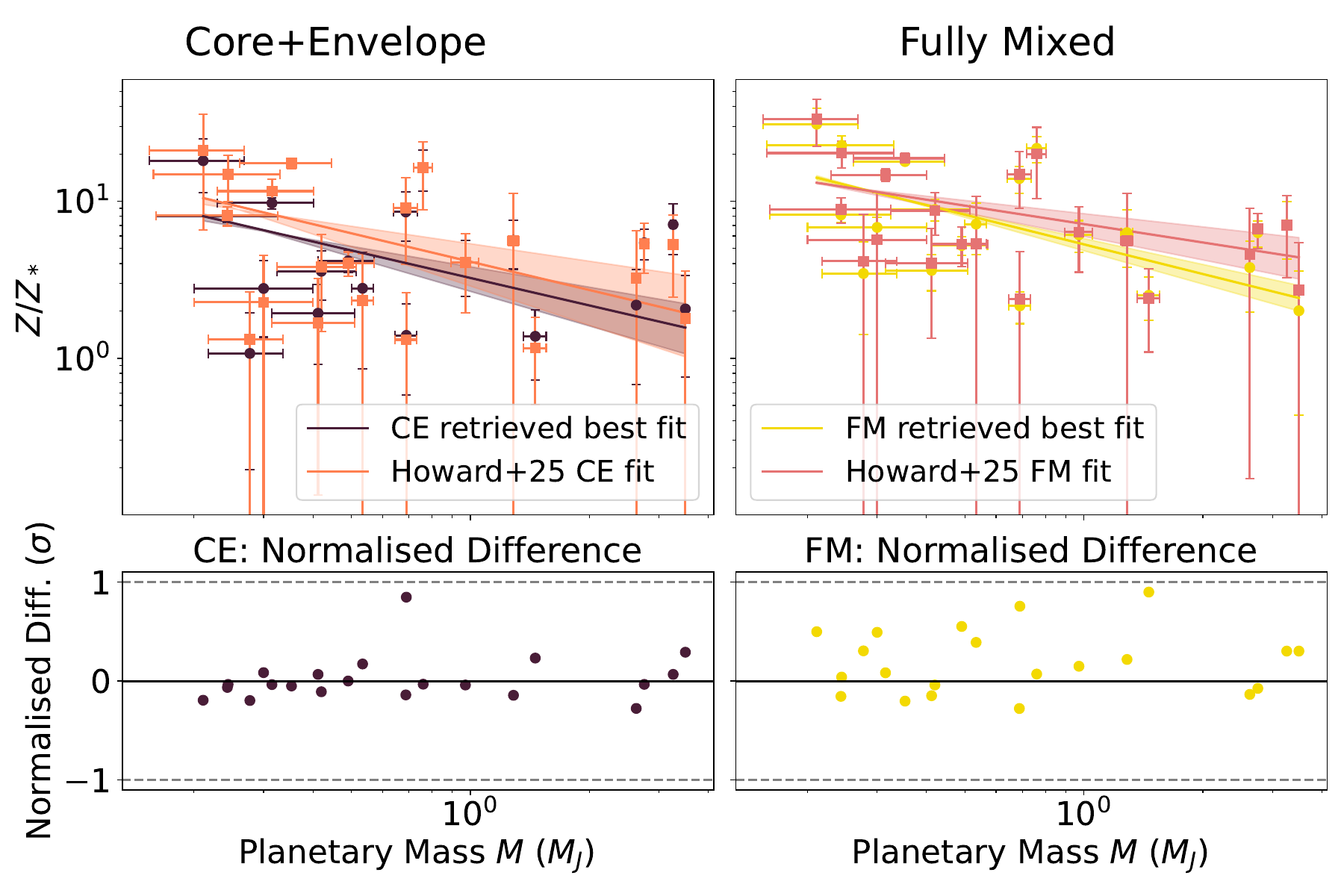}
    \caption{Output comparison between the data retrieved in this study, and the one found in \citet{Howard_2025}. Upper panels: Bulk metallicity (normalised by the star's metallicity) over mass for the initial 21 planets (as mentioned in Fig.~\ref{tab:sample_fullwidth_1} and \ref{tab:sample_fullwidth_2}) both for FM and CE structures. Bottom panels: Normalised difference (calculated with eq.~\ref{eq.16}) between the two estimates.}
    \label{fig.B1}
\end{figure}
In order to assess the strength and accuracy of the retrieved $Z$ for the sample used, we took advantage of the fact that the used dataset is the same as the one used by \citet{Howard_2025}. We thus took the initial sample of 21 exoplanets with the most precise data on Radius ($R_{err}$<3\%), and compared it with the same planets processed with the method explained in \citet{Howard_2025}. Results are shown in Fig. \ref{fig.B1}, where a visual comparison between the trends for the $Z$/$Z_{\star}$ versus $M$ is shown. To note how the best-fit lines vary from the ones retrieved with the full dataset of 44 exoplanets, showing the variability and unreliability of such curves for model prediction. Howard also discussed in their paper the shift of the slope $\alpha$ between the dataset, considering only $R_{err}$<3\%, or for the broader range considered of $R_{err}$<5\%. In the bottom panels of Fig. \ref{fig.B1}, it is shown how we assessed the difference between the data from this paper compared to the ones from \citet{Howard_2025}. While a simple subtraction could provide information about the difference between the two mean values, we also wanted to investigate whether each ranges of values (so accounting for data uncertainty) lie within the other. To do so, a normalisation using the standard deviations of the data has been computed. The normalised difference can be expressed with the following formula:
\begin{equation}
\label{eq.16}
    \sigma = \frac{\text{$CE/FM_{Howard}$} - \text{$CE/FM$}}{\sqrt{\sigma_{\text{Howard}}^{2} + \sigma_{\text{CE/FM}}^{2}}} .
\end{equation}

Results regarding this comparison were considered positive. While there is a slight change in the predictive trends between the two structures CE and FM (slight shift in $\beta$ for CE, and in $\alpha$ for FM compared to the ones found in Howard's. The results for the normalised difference ultimately confirmed the similarities between the two retrieved results, and their respective methods; All values lie between a standard deviation of 1 and -1 (or |$\sigma$|< 1), showing statistical consistency for all 21 cases compared. Furthermore, the normalised difference does not show any systematic error related to the mass (since there is no trend in the normalised difference $\sigma$). We thus conclude by saying that there is no discouraging difference between the two results (as the difference between the two approaches can be explained to be within the data error range and thus not due to randomness). We then accepted the validity of our approach, explained in Section 2.3, and proceeded with the inspection of the additional structures.

\section{Model assumption-driven changes}
\label{app:modass}
To isolate the impact of structural assumptions on the inferred metallicity, we quantified model-dependent offsets in $Z$ across the four models (FM, CE, DCA, DCA3) by forming differences relative to FM, namely $\Delta Z_{\mathrm{FM\text{-}CE}}\equiv Z_{\mathrm{FM}}-Z_{\mathrm{CE}}$, $\Delta Z_{\mathrm{FM\text{-}DCA}}\equiv Z_{\mathrm{FM}}-Z_{\mathrm{DCA}}$, and $\Delta Z_{\mathrm{FM\text{-}DCA3}}\equiv Z_{\mathrm{FM}}-Z_{\mathrm{DCA3}}$. 
\begin{figure}[b]
    \centering
    \includegraphics[width=0.5\textwidth]{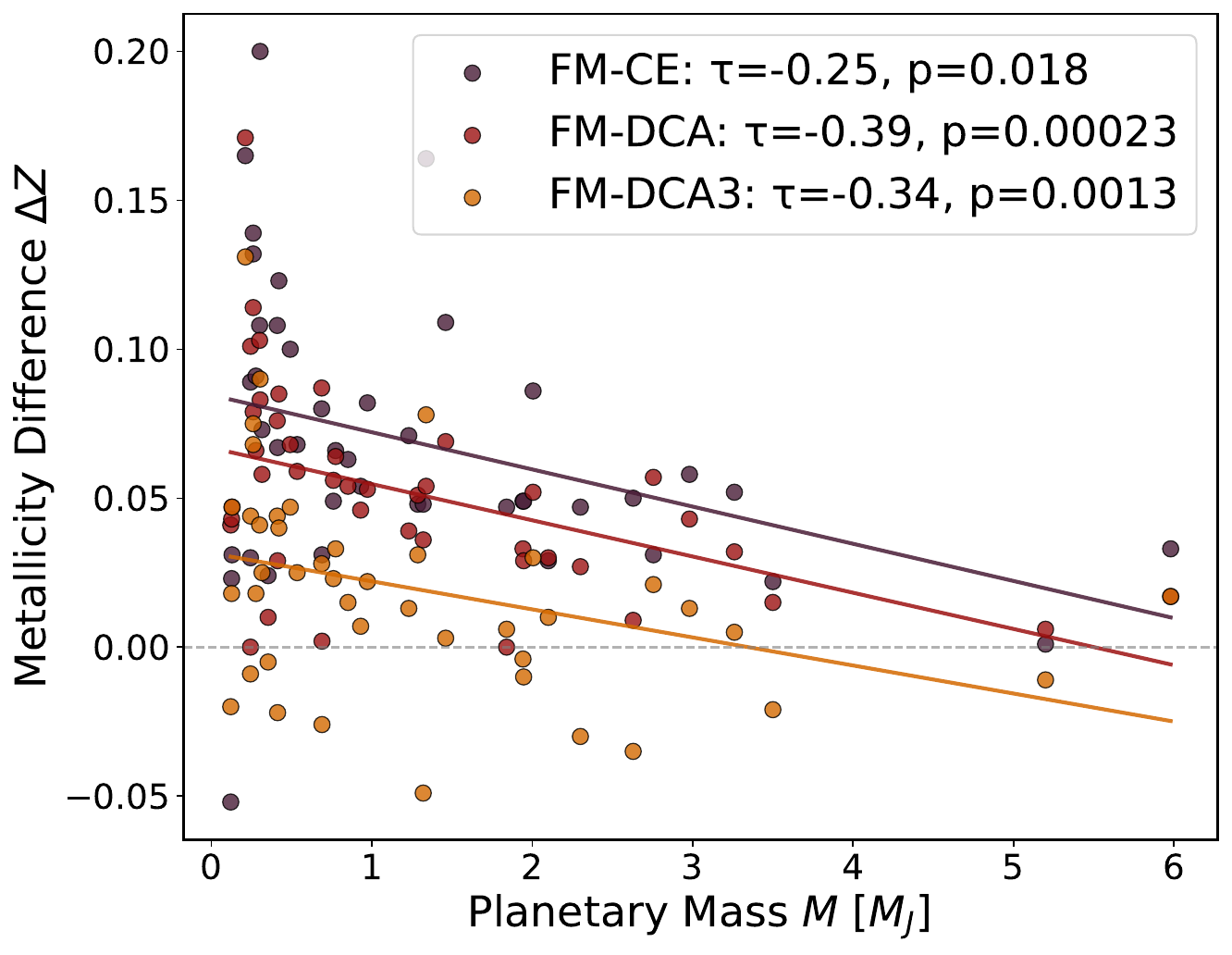}
    \caption{Change in bulk metallicity according to the assumed structure and their difference relative to the FM model over mass.}
    \label{fig.b2}
\end{figure}
Figure \ref{fig.b2} aims at showing such changes in the assumed structure over the full mass regime of the inspected exoplanets. The three lines show the best-fit lines of the data that is computed by the subtraction between the two mean points according to the models compared. No $\sigma$ in the models has been accounted for, as only the mean points have been inspected. For this reason, no error protraction has been shown on the best-fit lines, which have been kept only for visual reasons. Nonetheless, without counting model errors, the $\Delta$$Z$ for all three models being compared to FM show statistical validity as shown by the Kendall's $\tau$ results (and respective p-values) shown on the Figure's legend. We can see that, on average, all the models tend to reach 0 (No $Z$ difference between two structurally different models) in the high-$M$ regime. While the best-fit line for $\Delta$$Z_{FM-CE}$ approaches zero, the difference between the FM-DCA and FM-DCA3 goes to the negative (meaning that FM is no longer the structural end-member, but DCA and DCA3 overpredict $Z$ more than FM). This is caused by the forcing of respectively stellar and three times stellar abundances in their structures. Low-$Z$ scenarios (usually in the high-$M$ region) will be 'ruled out' by forcing an enriched metallicity in the atmosphere. This could suggest that low-$Z$ planets are unlikely to be super-enriched in heavy elements in their atmospheres. This agrees with the theory exposed in \citet{Fortney_2013}. Further studies must be made to investigate if that is the case, or if degeneracies between mass distribution in massive planets might overshadow the overall magnitude of the opacity effect proved to be dominant by this study.

\section{WLS $\chi^2$ test}
\label{app:wsl}
To assess the reliability of the power-law fits, we performed a WLS $\chi^2$ test, which accounts for individual data uncertainties rather than relying solely on $R^2$. In WLS, each residual is scaled by its measurement error:
\begin{equation}
\chi^2_{\rm obs} = \sum_{i=1}^{N} \left(\frac{y_i - f(x_i;\theta)}{\sigma_i}\right)^2,
\label{eq:wsl}
\end{equation}

where $y_i$ are the data, $f(x_i;\theta)$ the model, and $\sigma_i$ the 1$\sigma$ uncertainties. Under the null hypothesis that the model describes the data and errors are Gaussian, $\chi^2_{\rm obs}$ follows a chi-square distribution with $\nu = N - p$ degrees of freedom.

For $N=43$ and $p=2$, $\nu=41$, giving $\chi^2_{(0.05;\,41)} \approx 56.9$. Fits are accepted at the 5\% level if $\chi^2_{\rm obs} \le \chi^2_{(0.05;\,\nu)}$. In all cases, $\chi^2_{\rm obs}$ exceeds this threshold, indicating that the fits do not reproduce the data within the quoted errors. For the $Z$–$M$ relation, the observed values are:
\[
\chi^2_{\rm obs} = 580.4\ {\rm (CE)},\ 465.4\ {\rm (DCA)},\ 311.7\ {\rm (DCA3)},\ 243.7\ {\rm (FM)},
\]
corresponding to reduced $\chi^2_{\rm red}$ values of 14.2, 11.4, 7.6, and 5.9, respectively. Since $\chi^2_{\rm red} \gg 1$, we reject the null hypothesis and conclude that the observed scatter exceeds that expected from measurement errors alone. The diversity of planetary metallicities likely reflects intrinsic astrophysical variability rather than observational noise.

\end{appendix}

\end{document}